\documentclass[reqno,12pt]{amsart}
\usepackage{amsmath, latexsym, amsfonts, amssymb, amsthm, amscd}

\usepackage{xcolor,graphicx}

\usepackage{hyperref}

\usepackage{ucs}
\usepackage[applemac]{inputenc}
\fontfamily{Garamond} 
\advance\hoffset -.75cm                          

\usepackage[normalem]{ulem}

\numberwithin{equation}{section}

\newtheorem{Theorem}{Theorem}[section]

\newtheorem{Remark}[Theorem]{Remark}

\DeclareMathSymbol{\leqslant}{\mathalpha}{AMSa}{"36} 
\DeclareMathSymbol{\geqslant}{\mathalpha}{AMSa}{"3E} 
\DeclareMathSymbol{\eset}{\mathalpha}{AMSb}{"3F}     
\renewcommand{\leq}{\;\leqslant\;}                   
\renewcommand{\geq}{\;\geqslant\;}                   

\newcommand{\bra}{\langle}
\newcommand{\ket}{\rangle}

\newcommand{\cG}{\ensuremath{\mathcal G}}

\newcommand{\bbB}{{\ensuremath{\mathbb B}} }

\newcommand{\bbD}{{\ensuremath{\mathbb D}} }

\newcommand{\bbP}{{\ensuremath{\mathbb P}} }

\newcommand{\bbR}{{\ensuremath{\mathbb R}} }

\newcommand{\gp}{\varphi}

\title{
System driven out-of equilibrium by weak 
 contacts with reservoirs.
}

\author[Bodineau]{Thierry Bodineau$^\dagger$}
\address[$\dagger$]{CNRS, I.H.E.S., 35 Route de Chartres, 91440 Bures-sur-Yvette, France}

\author[Derrida]{Bernard Derrida$^\clubsuit$}
\address[$\clubsuit$]{Coll\`ege de France, 11 place Marcelin Berthelot, 75005 Paris, France}
\address[$\clubsuit$]{Laboratoire de Physique de l'Ecole Normale Sup\'erieure, ENS, Universit\'e PSL, CNRS, Sorbonne Universit\' e, Universit\' e de Paris, F-75005 Paris, France}

\thanks{{\it Acknowledgments} : At an early stage of this work, we have benefited of the hospitality of the Newton Institute in Cambridge during the 2024 program New Statistical Physics in Living Matter.}

\begin{document}

\maketitle

%
%

\begin{center}
{\it This paper is dedicated to Claudio Landim, on the occasion of his $60^{th}$ birthday, 
for his long-standing contributions  to the theory of  particle systems.
}
\end{center}

\begin{abstract}
The non-equilibrium behavior of particle systems driven by reservoirs has been extensively studied in recent years. In one dimension, various regimes have been explored depending on the coupling strength to the reservoirs. 
 In this paper, we investigate the role of the dimension and of the geometry of the contacts with the reservoirs.
For the symmetric simple exclusion process with point contact reservoirs, we show that in dimension 2, as in one dimension, three  different regimes occur depending on the coupling strength. On the other hand in dimensions 3 and higher, there  exists only a weak coupling regime which is very sensitive to the microscopic structure of the contacts.
We then argue that for reservoirs with mesoscopic size contacts the macroscopic fluctuation theory remains in force and
we propose an extension of the additivity principle for multiple mesoscopic reservoirs.
\end{abstract}

\section{Introduction}

Systems driven out of equilibrium by  contacts with several reservoirs (such as heat baths or particle reservoirs) 
 are a central topic  in statistical mechanics. 
Such driving generates a flux through the system, typically resulting in steady-state distributions that exhibit long-range correlations absent at equilibrium.
Lattice gas models evolving according to stochastic dynamics provide a well defined 
example of such systems
 (see \cite{Kipnis-Landim,BDGJL3,BDGJL6,Derrida 2007} in particular for an overview  
    on the way to describe lattice gases  by the macroscopic fluctuation theory).
In this paper, we focus on the effect of the contacts between the reservoirs and the bulk  to analyse the influence of their geometry  on the current flowing through the system.

To fix ideas,  we will consider   the concrete case of the symmetric simple exclusion process (SSEP) on 
a general finite graph $\Lambda$,  and  in particular the case of a cube $\Lambda = \{1, \dots , L\}^d$. The symmetric simple exclusion process on $\Lambda$ is a Markov chain with configurations $\eta(t) =\{ \eta_i (t) \}_{i \in \Lambda}$ where $\eta_i$  takes values $\{0,1\}$ indicating whether  site $i$ is empty or occupied.
At  exponential times, each particle attempts to jump uniformly   to one of its   neighbors on the graph. 
If the target site is already occupied, the jump is cancelled so that the exclusion condition of at most one particle per site remains satisfied. These dynamical rules of the SSEP are  reversible with respect to  any uniform measure on $\{0,1\}^\Lambda$ with a fixed number of particles. 
Given two sites $a, b$ in $\Lambda$,  one can add source terms which are injecting and removing particles in $a, b$ at some fixed rates :
\begin{equation}
\label{eq: rates definition}
\begin{aligned}
\text{at site $a$ : particles are injected at rate $\alpha$ and removed at rate $\gamma$,}\\
\text{at site $b$ : particles are injected at rate $\delta$ and removed at rate $\beta$.}
\end{aligned}
\end{equation}
Different rates at sites $a, b$  (more precisely if ${\alpha\over \gamma} \neq {\delta \over \beta} $) generate a flux which  drives  the system out of equilibrium.

\medskip

For a one-dimensional system with $\Lambda = \{1, \dots , L\}$, it is natural to choose the reservoirs at the extremities $a=1$, $b=L$.
In the large $L$ limit, after an appropriate rescaling, the density of SSEP evolves according to the heat equation in the bulk and the strength of the reservoirs  determines the boundary conditions. 
More precisely,  rescaling the rates \eqref{eq: rates definition} as
$\tilde \alpha/L^\theta , \tilde \beta/L^\theta, \tilde \gamma/L^\theta, \tilde \delta/L^\theta$,  
 then there are 3 regimes depending on the parameter $\theta$ :
\begin{itemize}
\item $0 \leq \theta < 1$, the rates are fast enough to fix the macroscopic densities at the boundaries. The macroscopic limit corresponds to the heat equation with  these fixed   boundary conditions. This case has been studied for a wide variety of  diffusive systems  (see e.g.  \cite{BDGJL6,Derrida 2007} and references therein).
\item $\theta = 1$ is critical in the sense that the densities at the reservoirs are coupled to the bulk evolution according to the Robin conditions. The hydrodynamic limit has been studied in \cite{Baldasso-Menezes-Neumann-Souza,Goncalves}  as well as the fluctuations in \cite{Goncalves-Jara-Menezes-Neumann}
and the large deviations of the current in \cite{DHS,Franco-Goncalves-Landim-Neumann, Bouley-Landim, BEL, Saha-Sadhu}.
\item $\theta > 1$, the effects of the reservoirs  evolve on a time  scale slower than the bulk diffusion, so that the hydrodynamic limit is given by the heat equation with Neumann boundary conditions
(see \cite{Erignoux-Goncalves-Nahum,Franco-Goncalves-Neumann, Landim-Velasco}).
\end{itemize}
These three regimes can be understood in terms of the current. If the densities $\rho_a,\rho_b$ at the boundaries are different, a current of order $|\rho_a - \rho_b|/L$ flows through the system, in agreement with Fick's law. Thus if the reservoir rates are faster than $1/L$ the boundary densities are determined by the reservoirs.  On the other hand the critical case corresponds to a balance between injection rates and the flux in the bulk.
As the structure of the steady states is determined by the transport properties, the three regimes above  lead to different types of steady states.

Let us note that other types of reservoirs have been also studied for one-dimensional models, including non reversible mechanisms \cite{Erignoux-Landim-Xu,Erignoux-2018,Landim-Mangi-Salvador}
and different injection mechanisms leading to unexpected phenomena  \cite{De Masi-Presutti-Tsagkarogiannis-Vares, Colangeli-De Masi-Presutti 1, Colangeli-De Masi-Presutti 2,De Masi-Merola-Presutti}.

\medskip

For systems in contact with reservoirs, the current is a key parameter to describe the non-equilibrium behaviors. 
The statistics of the current have been extensively studied for diffusive dynamics, in particular by computing the large deviations \cite{Derrida-doucot-Roche,Bodineau-Derrida-2004-PRL,BDGJL-2005-PRL,BDGJL-2006-JSP,Bodineau-Derrida-2006-JSP, Bodineau-Derrida-2007,Appert-Rolland_Derrida_Lecomte_Van Wijland_(2008),Lecomte-Imparato-vanWijland, Espigares_Hurtado-Garrido-PRE-(2016), Hurtado-(2025)}.
For general diffusive dynamics, the current large deviations are determined by a space-time variational problem
\cite{BDGJL-2005-PRL,BDGJL-2006-JSP,Bodineau-Derrida-2005-Phys-rev}
involving only macroscopic observables and  two transport coefficients: the diffusion $D(\rho)$ and the mobility $\sigma(\rho)$, both indexed by the local density $\rho$.
For small current deviations, one expects  that this complicated time dependent variational problem reduces to a much  simpler time independent variational problem  (called the additivity principle  \cite{Bodineau-Derrida-2004-PRL, Bodineau-Derrida-2007, Saito-Dhar, Hurtado-Garrido-PRL-(2009),Hurtado-Garrido-PRE-(2010),Hurtado-(2025)})
which allows to determine all   the cumulants of the current. 
For  larger  deviations of the currents, dynamical phase transitions have been observed, for some dynamics,  leading to time dependent optimisers to achieve the large deviation conditioning.
The phase transitions to a time dependent regime can occur in many ways and there is no complete description of the different mechanisms, see e.g.  \cite{Bodineau-Derrida-2005-Phys-rev, Baek-Kafri-Lecomte-2, Shpielberg-Yaroslav-Akkermans, Hurtado-Garrido-PRL-(2009), Hurtado-Garrido-PRE-(2010), Hurtado-Garrido-PRL-(2011),Espigares_Hurtado-Garrido-PRL-(2013), Zarfaty}. 
In contrast,  when the time independent variational problem holds   (i.e. the additivity principle), the hydrodynamic description predicts a universal structure of the current large deviations which is independent of the dimension, the geometry of domains or the location of the reservoirs \cite{Akkermans-Bodineau-Derrida-Shpielberg}.
More precisely, it was shown that all the cumulant ratios of diffusive systems  were universal in the sense that  they 
 depend only on the transport coefficients   $D(\rho)$ and $\sigma(\rho)$ but  not  on the  space dimension  $d$ or on the geometry of the contacts. This universality prediction was confirmed in \cite{Akkermans-Bodineau-Derrida-Shpielberg} by numerical calculations in $d=2$ but not in $d=3$. 
We believe that this discrepancy (in $d=3$) is due to the fact that the systems had point contacts with the reservoirs. 
  In this case,  the current as well as all its fluctuations are dominated by the immediate  vicinity of the contacts,  a region which is out of reach of the macroscopic fluctuation theory.

\medskip

To understand the numerical discrepancy observed in \cite{Akkermans-Bodineau-Derrida-Shpielberg}, we  study in this paper the sensitivity of the current to the geometry of the reservoir contacts. Depending on the dimension and on the domain, we will show that point contacts may lead to different behaviors.  
Note that understanding the connection between different scales is a challenging problem often encountered in modeling.
For example, crowd motion can be well approximated on large scales by fluid dynamics but this description is no longer valid to take into account a narrow passage (as a door) and matching both scales requires a very specific analysis, see e.g. \cite[Figures 18, 19]{maury2011}. The question raised in this paper regarding small reservoirs is of the same nature.
The main results are summarized below. 

\medskip

In Section \ref{sec: Microscopic computations}, we compute  the mean current of the SSEP  in a domain $\Lambda = \{1, \dots , L\}^d$ for microscopic point contacts  as in 
(\ref{eq: rates definition}).  We show that different behaviors occur depending on the recurrence  or the transience of the simple random walk. In dimension 2, 
the mean current decays as $1/\log L$ and for rates rescaled by $1/ (\log L)^\theta$, we  identify 3 regimes as in the one-dimensional case. In dimensions $d \geq 3$, the current is always of order $1$ and the statistics of the current will  depend strongly on the microscopic fluctuations  in a neighborhood of each reservoir. 
This explains the numerical discrepancy in \cite{Akkermans-Bodineau-Derrida-Shpielberg} previously mentioned.
As, in dimension $d \geq 3$,  the current is  always limited by the  flux through these point contacts,  there is only the regime of weak contacts. 

For mesoscopic reservoirs, i.e. with contacts of intermediate sizes $\ell$ (with $1 \ll \ell \ll L$), we expect that the macroscopic description of the large deviations still applies  and that the universality results predicted in 
\cite{Akkermans-Bodineau-Derrida-Shpielberg} remain valid, even though the average flux as well as all the cumulants of the current are limited by the size of the contacts (i.e. are of order $\ell^{d-2}$ {  in $d \geq 3$)}
(see Section \ref{subsec: Universal behavior for mesoscopic reservoirs}).
In dimensions larger than $3$, the influence of mesoscopic reservoirs is localised so that distant reservoirs interact very weakly. For this reason, the additivity principle \cite{Bodineau-Derrida-2004-PRL} can be generalised to compute the current large deviations for multiple currents coming from several mesoscopic reservoirs (see Section \ref{subsec: Extension of the additivity principle}).


\section{Microscopic computations for the mean current}
\label{sec: Microscopic computations}

In this section, we obtain the mean current of the SSEP for different  graphs in contact with reservoirs. We define  the microscopic current during the time interval $[0,t]$ from a reservoir in contact with site $a$  by
\begin{align}
\label{eq: microscopic current}
Q^{(a)}_t = & \text{\it Number of particles created at $a$ during $[0,t]$}  \\
& -  \text{\it Number of particles annihilated at $a$ during $[0,t]$}. \nonumber
\end{align}
In the steady state, we  denote the corresponding mean current as $J_a = \langle Q^{(a)}_t \rangle / t$ (which is independent of $t>0$).

\subsection{Mean current on a general graph}

For a general graph $\Lambda$,  one can write explicit  formulas for this  steady state  current $J_a$.
To be specific, let us first consider the case with 2 reservoirs connected to sites $a,b$ as in \eqref{eq: rates definition}. 
If  $J_a$ (resp $J_b$)  is  the mean current flowing from site $a$ (resp $b$)  into the bulk and  if the particle cannot accumulate in  the bulk one has  $J = J_a = - J_b$.
By Kirchhoff's law, the steady state profile is solution of a discrete Poisson equation with source terms at $a,b$
\begin{equation}
\forall x \in \Lambda, \qquad 
\sum_{y \, \sim \, x} \langle \eta_y \rangle  - \langle \eta_x\rangle  = -  J(\delta_{x,a} - \delta_{x,b}) ,
\label{eq: Laplace 2 sites} 
\end{equation}
where $\langle \cdot \rangle$ stands for the expectation under the steady state and the summation is over the neighboring sites $y$ of $x$ in the graph $\Lambda$.
Furthermore (see  \eqref{eq: rates definition})  the steady state current through the system from contact $a$ to contact $b$ is given by
\begin{equation}
J= \alpha -(\alpha + \gamma) \langle \eta_a \rangle = (\beta+ \delta) \langle \eta_b \rangle - \delta  
\label{eq: boundary conditions 2 sites 0} 
\end{equation}
which expresses that, in the steady state, the particles do not accumulate.
It is convenient to rewrite the steady state current with the following notations 
\begin{equation}
\rho_a = \frac{\alpha}{\alpha + \gamma}, \quad 
A= \frac{1}{\alpha + \gamma}, \quad 
\rho_b =  \frac{\delta}{\beta+ \delta}, \quad 
B = \frac{1}{\beta+ \delta},
\label{eq: reservoirs notations} 
\end{equation}
so that \eqref{eq: boundary conditions 2 sites 0}  gives
\begin{equation}
\langle \eta_a \rangle = \rho_a - A J
\quad \text{and} \quad 
\langle \eta_b \rangle = \rho_b + B J.
\label{eq: boundary conditions 2 sites} 
\end{equation}
$\rho_a$ and $\rho_b$ represent the densities of the reservoirs whereas $A$ and $B$ are related to the strengths of the contacts with the reservoirs: small $A$ or $B$ represent strong contacts while  large values of $A$ or $B$ correspond to weak contacts.

\medskip

For any site $c$ in $\Lambda$,  
 the  Green function is   defined as the  solution of
\begin{equation}
\forall x \in \Lambda, \qquad 
\sum_{y \, \sim \,  x} \big( G^{(c)}_y  - G^{(c)}_x\big)  +   \delta_{x,c} = \frac{1}{|\Lambda|}
\label{eq: Green function} 
\end{equation}
where $|\Lambda|$ stands for the number of sites in $\Lambda$. 
For a connected graph $\Lambda$, there is a unique solution $G_x^{(c)}$  up to an arbitrary additive constant independent of $x$.
Explicit expressions, as well as some numerical values and asymptotics,  of these Green functions  for cubic lattices with periodic or free boundary conditions  are given  in the appendix \ref{appendix: Green functions}. Also, it is well known that the Green functions are related to the return probabilities  \eqref{eq: ra return}
(see Appendix \ref{Appendix: return probability}).

The solution of the Poisson equation \eqref{eq: Laplace 2 sites} can then  obtained by a superposition of two Green functions
\begin{equation}
\forall x \in \Lambda, \qquad 
\langle \eta_x \rangle 
= J \big( G^{(a)}_x  - G^{(b)}_x \big) + K ,
\label{eq: solution 2 sources} 
\end{equation}
where the parameters $J,K$ are determined by 
\begin{equation}
 \langle \eta_a \rangle = J (G^{(a)}_a - G^{(b)}_a) +  K 
%
\quad \text{and} \quad 
 \langle \eta_b \rangle = J (G^{(a)}_b - G^{(b)}_b) +  K 
\label{eq: contraintes bords 2 sources} 
 \end{equation}
and the 2 equations in \eqref{eq: boundary conditions 2 sites} prescribing the boundary conditions.
This gives 
\begin{equation}
J=  \frac{\rho_a-\rho_b}
{A+B + G^{(a)}_a + G^{(b)}_b - G^{(a)}_b - G^{(b)}_a } ,
\label{eq: solution J,K 2 sources} 
\end{equation}
and
\begin{equation}
\langle \eta_x \rangle= \rho_b +\frac{(B + G_x^{(a)} - G_x^{(b)} - G_b^{(a)} + G_b^{(b)} )\,  (\rho_a-\rho_b)} 
 {A+B + G^{(a)}_a + G^{(b)}_b - G^{(a)}_b - G^{(b)}_a } .
\label{etax}
\end{equation}

\medskip

The above computation can be generalised to an arbitrary number of sources $\{ i_1, \dots, i_k \}$
with currents $\{ J_1, \dots, J_k \}$ exiting from each source and satisfying the particle conservation constraint $J_1 +  \dots + J_k =0$.
The conservation law \eqref{eq: Laplace 2 sites}  becomes 
\begin{equation}
\forall x \in \Lambda, \qquad 
\sum_{y \, \sim \, x} \langle \eta_y \rangle  - \langle \eta_x\rangle  = - \sum_{\ell = 1}^k  J_\ell \,  \delta_{x,i_\ell}  
\end{equation}
and the boundary conditions \eqref{eq: boundary conditions 2 sites} at each   reservoir becomes
\begin{equation}
\ell \leq k, \qquad 
\langle \eta_{i_\ell} \rangle = \rho_{i_\ell} - A_\ell \, J_\ell,
\label{eq: boundary conditions k sites} 
\end{equation}
where  we extrapolated the notations \eqref{eq: reservoirs notations}.

As in \eqref{eq: solution 2 sources},  the steady state is obtained by superpositions of  Green functions
\begin{equation}
\forall x \in \Lambda, \qquad 
\langle \eta_x \rangle =   \sum_{\ell = 1}^k  J_\ell \, G^{(i_\ell)}_x + K .
\label{eq: solution k sources} 
\end{equation}
The  currents $\{ J_1, \dots, J_k \}$ and the parameter $K$ are then determined by the $k$   constraints  \eqref{eq: boundary conditions k sites}
\begin{equation}
u \leq k, \qquad 
\langle \eta_{i_u}  \rangle =  \sum_{\ell = 1}^k  J_\ell  \,  G^{(i_\ell)}_{i_u} + K 
 =\rho_{i_u} - A_u J_u.
\label{eq: contraintes bords k sources} 
 \end{equation}
and the conservation law $J_1 + J_2 + \dots + J_k =0$.
This is a system of linear  equations allowing to find expressions of the $J_\ell$ and $K $ as ratios of determinants which generalize \eqref {eq: solution J,K 2 sources}.


\subsection{Effect of the contacts for different dimensions}

From \eqref{eq: solution J,K 2 sources}, the mean current for the SSEP is determined by the Green function solution of \eqref{eq: Green function}.
We are now going to discuss its asymptotics for several geometries and show that 
different regimes are related to the recurrence/transience of the simple random walk on the corresponding graphs.

\subsubsection{One-dimensional SSEP}
 
For an open system on the domain $\Lambda = \{ 1, \dots ,L\}$ with reservoirs at $a =1$ and $b= L$, the Green functions  solution of \eqref{eq: Green function} are given by 
(\ref{G-open-1}) in Appendix \ref{appendix: Green functions}.
This leads to 
\begin{equation}
J=  \frac{\rho_a-\rho_b}{A+B + L-1 } ,
\label{eq: solution J 2 sources d=1} 
\end{equation}
 where $A,B$ are defined in \eqref{eq: reservoirs notations}.
The linear growth  of the Green function with $L$ implies that the current through the system is sensitive  to the reservoir rates only if $A,B$ are of order $L$ or larger, i.e. if the rates $\alpha + \gamma$ and $\beta + \delta$ are weak. 
This leads to the three regimes mentioned in the introduction
 with  transition at the scale $1/L$ (or $\theta =1$).
The divergence of the Green function as $L \to \infty$ is related to the recurrence of the simple random walk (see Appendix \ref{Appendix: return probability}).

In particular in the critical case (see e.g. \cite{Baldasso-Menezes-Neumann-Souza,Goncalves}), for large $L$  and weak contacts  given by
\begin{equation}
\label{weak-c}
\alpha= {\rho_a  \over L  \, \Gamma_a }, \quad   
\gamma= {(1-\rho_a)  \over L \, \Gamma_a }, \quad   
\beta= {(1-\rho_b)  \over L \, \Gamma_b}, \quad  
\delta= {\rho_b  \over L \, \Gamma_b},
 \end{equation}
 the current  \eqref{eq: solution J 2 sources d=1}  scales as
$$J\sim
{1 \over  1 + \Gamma_a+\Gamma_b} \, {\rho_a - \rho_b  \over L } .
$$

\subsubsection{Two-dimensional SSEP} 

 In dimensions 1 and 2, the random walk is recurrent and the Green function $G_c^{(c)}$ grows  with the system size (see Appendix \ref{appendix: Green functions}). 
 On the domain $\Lambda = \{1, \dots, L\}^2$, the growth of the Green function is  logarithmic in $L$  
  and, as in the one-dimensional case,
 one expects three different regimes depending on the strength of the contacts  measured by the parameters $A,B$ introduced in \eqref{eq: reservoirs notations} :  
\begin{itemize}
\item Strong contacts where $A$ and $B$ are much smaller than $\log L$. Then for the domain $\Lambda = \{1, \dots, L\}^2$ with periodic boundary conditions and contacts $a,b$ at distances of order $L$
the current \eqref{eq: solution J,K 2 sources}  scales as 
\begin{equation}
\label{strong periodic}
J  \simeq \pi  \, {\rho_a-\rho_b \over \log L},  
\end{equation}
where we used the asymptotic \eqref{G-per-2} of the periodic Green function.
For the  domain $\Lambda = \{1, \dots, L\}^2$ with  free boundary conditions and contacts $a,b$ at corners of  the square : \{$a=(1,1)$ and $b=(L,L)$\} or \{$a=(1,1)$ and $b=(1,L)$\} or \{$a=(2,2)$ and $b=(L,L)$\}, the current  \eqref{eq: solution J,K 2 sources}  decays as 
\begin{equation}
\label{strong free}
J  \simeq {\pi \over 4 } \,  {\rho_a-\rho_b \over \log L},
\end{equation}
where we used the asymptotics \eqref{G-open-2} of the Green function.
Notice that the current \eqref{strong free} doesn't depend on the precise position of the contacts, as long as the distance between the contacts is of order $L$ and they remain close to the corners.
We stress that, for free boundary conditions,
 the prefactor depends on whether the source is at a finite distance from a boundary of the square or close to a corner (see Remark \ref{rem: Green function dependence c}).
\item Critical contacts 
with $A=  \Gamma_a \, \log L$ and $B= \Gamma_b \, \log L$ lead to the following corrections of 
\eqref{strong periodic} and \eqref{strong free} :
\begin{align*} 
     J &  \simeq {\pi \over 1  + \pi \Gamma_a + \pi \Gamma_b} \,  {\rho_a-\rho_b \over \log L} \ \ \ \ & \text{for periodic boundary conditions}, \\ 
     J &  \simeq {\pi \over 4  + \pi \Gamma_a+\pi \Gamma_b} \,{\rho_a-\rho_b \over \log L} \ \ & 
     \text{for free boundary conditions.}
\end{align*}
\item Very weak contacts, i.e. if $A$  and  $B \gg \log L$, the current scales as
$$
J  \simeq {\rho_a-\rho_b \over A+B}.
$$
\end{itemize}

\begin{Remark}
\label{rem: Green function dependence c}
For a square of size $L$ with free boundary conditions, the Green function \eqref{G-open} scales as  $G_c^{(c)} \simeq  C_c \log L$ where the constants $C_c$ depend on the position of the site $c$ on the square
($C_c ={2  \over \pi}$ if $c$ is near a  corner, $C_c= {1 \over \pi}$ if $c$ is near the side of the square but far from the corners and $C_c={1 \over 2 \pi}$ if $c$ is at distance of order $L$ from the boundaries). 
On the other hand, $G_{c'}^{(c)} \ll \log L$ for pairs of points $c,c'$ at distances of order $L$. 
This simplifies the calculation of the currents  from \eqref{eq: contraintes bords k sources} and one gets in the case of strong contacts ($A+B \ll \log L$) an asymptotic of the form :
\begin{equation}
J_l= {1 \over G_{i_\ell}^{(i_\ell)} } \left[ \rho_{i_\ell}  
-{S_1 \over S_2} \right] ,
\label{Jl}
\end{equation}
 where  the sums $S_1,S_2$ are determined from \eqref{eq: contraintes bords k sources} by the condition $J_1 + J_2 + \dots + J_k =0$ :
$$
S_1= 
\sum_{u=1}^k {\rho_{i_u} \over G_{i_u}^{(i_u)}}, 
\qquad  
S_2=\sum_{u=1}^k {1 \over G_{i_u}^{(i_u)}}.
$$ 
This is a generalization of \eqref{strong free} to the case of $k$ sources.
\end{Remark}

\bigskip

\subsubsection{The SSEP in  $d \geq 3$ :}
In dimension 3, and higher, random walks on an infinite lattice are transient. This is due to the fact that, in the  large $L$ limit,  the Green functions $G_c^{(c)}$ have a finite limit  
whereas $G_{c'}^{(c)} \to 0$ when $c$ and $c'$ are at distances of order $L$.
The main consequence is that, in contrast to the cases $d=1$ or $d=2$, there is a single regime, {\it the regime of weak contacts} when the contacts are at distances of order $L$: in particular  even if the contacts with the reservoirs are very strong (i.e. $A_\ell=0$), for large $L$,  the currents do not depend on the system size $L$.

The equations \eqref{eq: contraintes bords k sources} for the currents in the case of $k$ contacts (at distances of order $L$ from each other) can be solved easily as each reservoir acts locally in $d \geq 3$. 
In the steady state, the density $\langle \eta_x \rangle$, computed in \eqref{eq: solution k sources},  is essentially  constant equal to $\rho_\text{bulk}$ at large enough distance from the contacts. Thus  the parameter $K$ in \eqref{eq: contraintes bords k sources}  coincides with the bulk density $\rho_\text{bulk}$. We then get from \eqref{eq: contraintes bords k sources} that 
\begin{equation}
\label{eq: }
J_\ell = {\rho_{i_\ell} -\rho_\text{bulk} \over A_l + G_{i_\ell}^{(i_\ell)}}.
\end{equation}
Note that  $\rho_\text{bulk}$ is determined from \eqref{eq: contraintes bords k sources} by the constraint that $\sum_{\ell=1}^kJ_\ell =0$. This means   that the current at each contact is the same as if there was a single contact in an infinite system with a density $\rho_\text{bulk}$ at infinity.

In the case of two very strong contacts ($A=B=0$) on sites $a$ and $b$  the current is given by
\begin{equation}
\label{eq: mean current 3D}
J= {\rho_a-\rho_b \over G_a^{(a)} + G_b^{(b)} },
\end{equation}
when the distance between $a$ and $b$ is large enough.
The value of the current is now sensitive to the  precise microscopic position of the contact with respect to the boundary. 
This can be seen from the numerical values of the Green functions given in Appendix 
\ref{appendix: Green functions} :  for an open cube when $\rho_a=1$ and $\rho_b=0$, we get
\begin{itemize}
\item for  $a=(1,1,1)$ and $b=(L,L,L)$ or for 
   $a=(1,1,1)$ and $b=(L,1,1)$, then  $J  \simeq .69437$;
\item for  $a=(1,1,1)$ and $b=(L-1,L-1,L-1)$, then $J  \simeq .88746$;
\item for  $a=(2,2,2)$ and $b=(L-1,L-1,L-1)$, then $J  \simeq 1.22932$.
\end{itemize}

\section{Higher cumulants and large deviations}
\label{sec: Higher cumulants and large deviations}

\subsection{The cumulants}

In Section \ref{sec: Microscopic computations}, we  gave  explicit expressions for the mean current in terms of Green functions for  SSEP in  contact with  an arbitrary number of reservoirs. The next step is to consider higher cumulants of the current. 
For the SSEP, the variance of the current $Q^{(a)}_t$ \eqref{eq: microscopic current}  can be  computed (see [Eq. (3.18)] in \cite{Bodineau-Derrida-Lebowitz-2008} or  Appendix \ref{Appendix: variance}) in terms of the steady state density $\langle \eta_x \rangle$ \eqref{eq: solution k sources}  and two point correlation functions $\langle \eta_x \eta_y \rangle$. The latter solve also a set of linear closed equations  for which it is difficult to obtain  explicit solutions in dimensions larger or equal to 2 if the reservoirs drive the system out-of equilibrium.
In   one dimension, for a system of length $L$ and reservoirs with weak contacts as in (\ref{weak-c}), the fluctuations and  the large deviations of the current  were found in \cite{DHS,Saha-Sadhu}.  
In this case, the variance of the current is
$$
\lim_{t \to \infty} 
\frac{\langle (Q^{(a)}_t)^2 \rangle - \langle Q^{(a)}_t \rangle^2}{t  } 
=  
{3 (\rho_a + \rho_b) - 2(\rho_a^2 + \rho_a \rho_b + \rho_b^2) \over 3 (1+ \Gamma_a+\Gamma_b)\, L}   +  {2 \over 3} {\Gamma_a^3 + \Gamma_b^3 \over (1  + \Gamma_a  + \Gamma_b)^4 \, L  } (\rho_a - \rho_b)^2.
$$
This remains valid for strong contacts by setting $\Gamma_a= \Gamma_b = 0$ \cite{Derrida-doucot-Roche}.

\medskip

Computing the cumulants boils down to taking successive derivatives of the moment generating function $\lambda \mapsto  \frac{1}{t} \log \bra \exp ( \lambda Q^{(a)}_t) \ket$ and then to identify the large time limit $t \to \infty$. Alternatively, one can consider the opposite limits, namely first $t \to \infty$ and then take the derivatives at $\lambda =0$. This second procedure is related to large deviations and 
formally, away from phase transitions, it should also provide all the cumulants.
The two procedures are equivalent for the SSEP on any finite graph, because the number of possible microscopic configurations is finite. 
This will also be checked  in the simple case of independent random walks in Section \ref{sec: Simple random walks}.

Large deviations of the current have been studied intensively for one dimensional systems and also for  reservoirs with macroscopic contacts in higher dimensions \cite{Bodineau-Derrida-2004-PRL,BDGJL-2005-PRL,BDGJL-2006-JSP, Bodineau-Derrida-2007, Hurtado-(2025)}.  
That way,
 the cumulants of the current can be  predicted from the large deviation functional and, 
in the case of SSEP, these predictions agreed  with the exact computations in \cite{Derrida-doucot-Roche}.
In Section \ref{subsec: Universal behavior for mesoscopic reservoirs}, 
we will consider the case of several reservoirs   
with contacts of intermediate sizes $\ell$ ($1 \ll \ell \ll L$)  in dimension 3 and higher.
In $d \ge 3$, such mesoscopic reservoirs act locally in a way analogous to  the case of very weak contacts in $d=1$. 
{ As a consequence, we will show in \eqref{eq: probability LD k mesoscopic} that if the  additivity principle holds then 
the large deviations of the current can be decomposed as a sum of simpler local costs associated with each reservoir.
}

\subsection{Simple random walks}
\label{sec: Simple random walks}

We first consider the case of non interacting particles for which  the large deviations can be explicitly computed in terms of the return probability.
Let $\Lambda$ be a finite graph and $a,b$ two sites in $\Lambda$ acting as reservoirs. 
Inside the domain $\Lambda$, the particles evolve as simple random walks and  the (integer) number of particles $n_i$ at site $i$ can be arbitrarily large.
The rates at the reservoirs are given by 
\begin{equation}
\label{eq: rates definition-bis}
\begin{aligned}
\text{at site $a$ : particles are injected at rate $\alpha$ and removed at rate $\gamma \,  n_a$,}\\
\text{at site $b$ : particles are injected at rate $\delta$ and removed at rate $\beta \, n_b$.}
\end{aligned}
\end{equation}
In contrast  to \eqref{eq: rates definition} the annihilation rates are now proportional to the number of particles $n_a, n_b$ at the reservoirs. 
As in \eqref{eq: Laplace 2 sites}, the steady state density obeys the Laplace equation
\begin{equation}
\forall x \in \Lambda, \qquad 
\sum_{y \, \sim \, x} \langle n_y \rangle  - \langle n_x\rangle  = -  J(\delta_{x,a} - \delta_{x,b}) ,
\end{equation}
with the boundary conditions given by 
\begin{equation}
J= \alpha -  \gamma \langle n_a \rangle = \beta \langle n_b \rangle - \delta .
\end{equation}
Following \eqref{eq: solution J,K 2 sources}, the mean current is given by  
\begin{equation}
J=  \frac{\rho_a-\rho_b}
{A+B + G^{(a)}_a + G^{(b)}_b - G^{(a)}_b - G^{(b)}_a }
\quad \text{with} \quad 
\rho_a= \frac{\alpha}{\gamma}, \    \rho_b= \frac{\delta}{\beta}, 
\  A= \frac{1}{\gamma}, \   B= \frac{1}{\beta}.
\label{eq: solution J,K 2 sources iid} 
\end{equation}

\medskip

Starting from the steady state, we are going to compute the large time asymptotic of the generating function
$\frac{1}{t} \log \bra \exp ( \lambda Q^{(a)}_t) \ket$ of the  current $Q^{(a)}_t$ 
\eqref{eq: microscopic current} flowing through $a$.
As the domain $\Lambda$ is finite, this is equivalent to studying the total current flowing through the system.
If this was not the case, then the number of particles in the bulk at the initial or final times would be at least of order $t$.
This can be neglected in the large time asymptotic. 
Indeed for independent particles, the steady state is an explicit product Poisson measure and observing a  total  number of particles proportional to $t$ in the system has a vanishing cost of order  $\exp( - t \log t)$ which is much smaller than the large deviation scale we are interested in.

A particle created at $a$ contributes to the current $Q^{(a)}_t$ if it doesn't exit the system by $a$.
As the particles are independent, exiting by $b$ has a probability $p_{a \to b}$ for each particle. Thus during the time interval $[0,t]$, the
reservoir at $a$ will inject particles according to a Poisson law with rate   $\alpha $ and a proportion 
$p_{a \to b}$ of these particles will contribute positively to the current. 
The same  mechanism occurs at reservoir $b$ which is injecting particles in the opposite direction.
These probabilities   are  computed in \eqref{random walk} in terms of the return probabilities 
$r_a, r_b$ \eqref{eq: ra return} :
\begin{equation}
\label{eq: return probability}
\begin{aligned}
p_{a \to b} & = \bbP \Big[ \text{a particle created at $a$ is annihilated at $b$} \Big]
= { \beta  \over \beta+\gamma + \beta \gamma /  [ (1-r_a) \, {\rm deg}(a)]} ,
\\
p_{b \to a} & = \bbP \Big[ \text{a particle created at $b$ is annihilated at $a$} \Big]
= { \gamma  \over \beta+\gamma + \beta \gamma / [  (1-r_b) \,  {\rm deg}(b)] }.
\end{aligned}
\end{equation}

For large times, the law of the current $Q^{(a)}_t$ is equivalent to the law of the 
difference $N^{(a)}_t - N^{(b)}_t$ of two independent Poisson processes 
 $N^{(a)}_t, N^{(b)}_t$  
with rates $\alpha p_{a \to b}$ and $\delta \, p_{b \to a}$.
This leads to an explicit form of the cumulant generating function
\begin{align}
\lim_{t \to \infty} \frac{1}{t} \log \bra \exp ( \lambda Q^{(a)}_t) \ket
& = \lim_{t \to \infty} \frac{1}{t} \log \bra \exp ( \lambda N^{(a)}_t) \ket
+ \frac{1}{t} \log \bra \exp ( - \lambda N^{(b)}_t) \ket  \nonumber \\
& = \alpha p_{a \to b} (e^\lambda -1) + \delta \, p_{b \to a} (e^{- \lambda} -1).
\label{eq: cumulant generating}
\end{align}
The large deviation function of the current can then be computed by taking the Legendre transform.

Note that the previous analogy with independent point processes is not new and has been used already in several papers, e.g. \cite{BDGJL-2006-JSP,Dhar-Saito-Derrida}.
As observed in \cite{BDGJL-2006-JSP}, formula \eqref{eq: cumulant generating} goes beyond the case of independent variables : the current large deviations for the Zero Range Process in contact with reservoirs are the same as the one of the independent particles as each particle, in the Zero Range Process,  performs a  simple random walk up to a time change.

\subsection{Universal behavior for mesoscopic reservoirs}
\label{subsec: Universal behavior for mesoscopic reservoirs}

For systems driven out of equilibrium by macroscopic reservoirs, the large deviations of the current were predicted by a variational principle \cite{Bodineau-Derrida-2004-PRL,BDGJL-2006-JSP}.
In this section, we apply these results to compute the large deviations associated with mesoscopic reservoirs, i.e. with contacts much smaller than the size of the system, but large enough so that the macroscopic description still holds. The following discussion remains at a heuristic level and the claims will not be mathematically proved.

\subsubsection{Macroscopic description}

\begin{figure}[htbp]
\begin{center}
\includegraphics[width=2in]{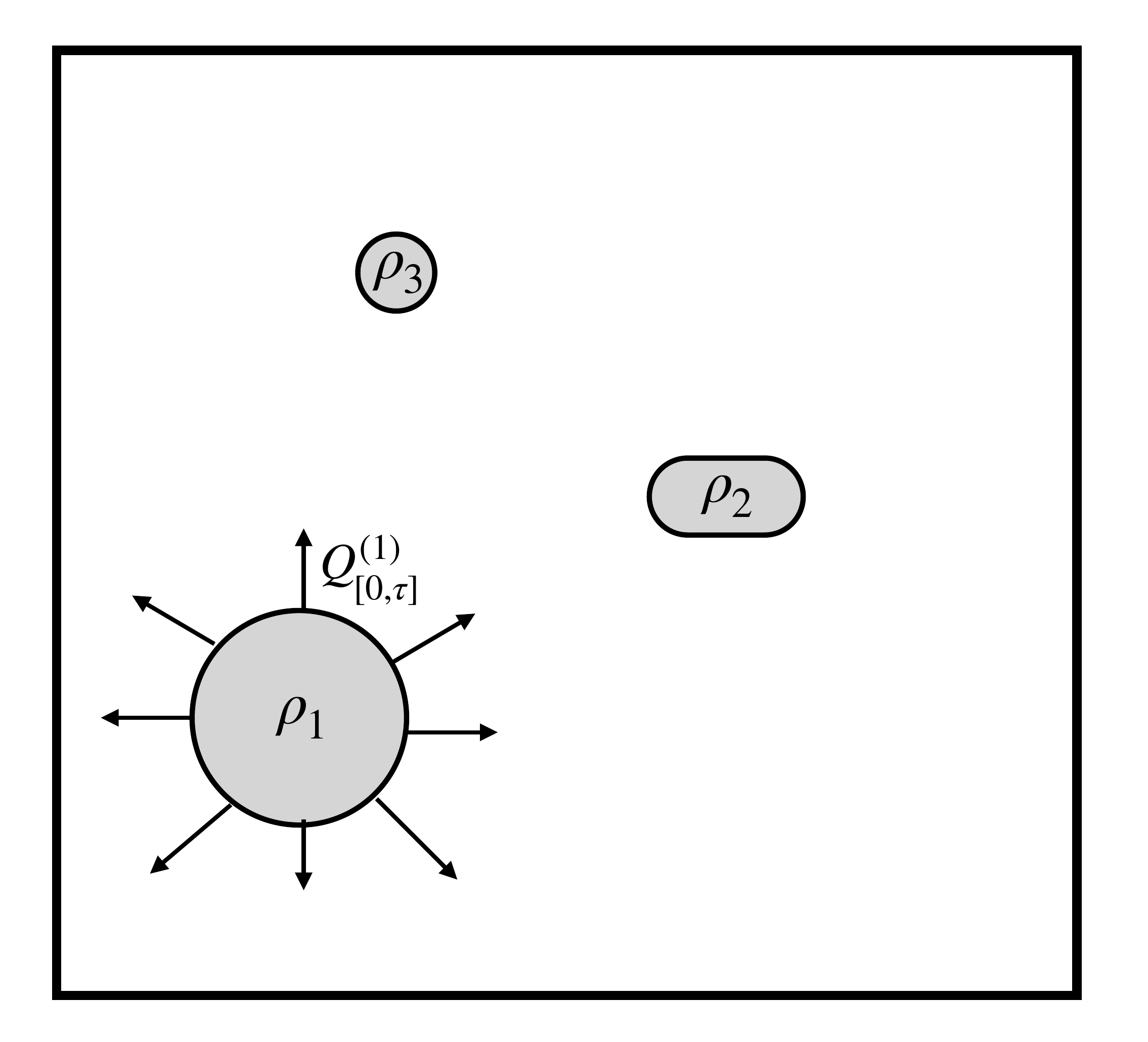} 
\caption{\small  In this figure, a  square domain $\bbD$ is depicted  with 3 holes  in gray representing 3 reservoirs at different densities $\rho_1, \rho_2, \rho_3$ and with different shapes. 
The flux $Q^{(1)}_{[0,\tau]}$ through the boundary $\partial \bbD_1$ of the first reservoir is represented by arrows.
On the boundary of the square, one could consider either Neumann boundary conditions (if there is no incoming flux) or another reservoir.}
\label{fig: multiple sources}
\end{center}
\end{figure}

We consider a microscopic diffusive dynamics in a domain $\bbD \subset \bbR^d$ discretized with a mesh $1/L$. 
On all the sites at the boundary $\partial \bbD$ of the domain, the density is fixed by reservoirs.
For concreteness, we assume  that the boundary $\partial \bbD$  is split  into $k$ disjoint sets $\partial \bbD_1, \dots, \partial \bbD_k$ with fixed density equal to the value $\rho_i$ on $\partial \bbD_i$ (see Figure \ref{fig: multiple sources}).
The macroscopic time $\tau$ corresponds to the microscopic time $t = \tau L^2$.
At the macroscopic scale, the stochastic dynamics is expected to be well approximated by the following fluctuating hydrodynamic equations 
\begin{equation}
\begin{aligned}
\label{eq: conservation law}
& \forall x \in \bbD, \qquad \qquad 
\partial_\tau \rho(\tau,x) = - \nabla \cdot q(\tau,x),\\
& \forall i \leq k, \forall x \in \partial \bbD_i, \qquad \rho(\tau,x) = \rho_i,
\end{aligned}
\end{equation}
and
\begin{align}
\label{eq: hydro limit}
q(\tau,x) = D \big( \rho(\tau,x) \big) \nabla \rho(\tau,x) + \frac{1}{\sqrt{L}}  \sqrt{\sigma \big( \rho(\tau,x) \big)}  \dot \zeta(\tau,x),
\end{align}
where $\rho \in \bbR^+$ is the density, $q   \in \bbR^d$ is the current and $\dot \zeta$ a space time white-noise. 
The diffusion coefficient $D$ and the mobility $\sigma$ are functions of the local density determined by the microscopic rates.

The macroscopic  current  during a macroscopic time interval $[0,\tau]$ is defined by integrating the flux $q(s,x)$ through a surface inside the domain. 
The total current flowing out of the region $\partial \bbD_i$ is denoted by 
$Q^{(i)}_\tau$ (see Figure \ref{fig: multiple sources}). 
This is the macroscopic  counterpart to the microscopic current \eqref{eq: microscopic current}.
We expect that in dimension $d$ the current through the macroscopic region $\partial \bbD_i$ 
 scales like $Q^{(i)}_\tau = O \big( L^{d-2} t \big)$ 
and $t = \tau L^2$ stands for the microscopic time.
In the following, we simply write $Q_{\tau} = \big( Q^{(i)}_\tau \big)_{i \leq k}$.

Given $u = (u_i)_{i \leq k}$ with $\sum_{i =1}^k u_i =0$, we want to evaluate the probability that 
the currents $Q_{\tau} = \big( Q^{(i)}_\tau \big)_{i \leq k}$ are proportional to $L^d \tau u$ in the large time limit. Note that the constraint $\sum_{i =1}^k u_i =0$ prevents the accumulation of particles inside the bulk. 
The fluctuating hydrodynamic equations \eqref{eq: hydro limit} lead to a natural guess for the large deviations in the limit $L \to \infty$ and then $\tau \to \infty$
\begin{align}
\label{eq: probability LD}
\bbP \Big[ \frac{Q_{\tau}}{L^d \tau} = u \Big] 
\simeq \exp \Big( - L^d \tau \; \cG_{\bbD} (u) \Big) ,
\end{align}
with 
\begin{align}
\label{eq: time dependent LD}
\cG_{\bbD} (u) 
= \lim_{T \to \infty} \inf_{q,\rho} \Bigg \{ \frac{1}{T} \int_0^T ds \int_\bbD dx 
\frac{ \big( q(s,x) + D \big( \rho(s,x) \big) \nabla \rho(s,x) \big)^2 }{2 \,   \sigma \big( \rho(s,x) \big)} 
\Bigg\},
\end{align}
where the infimum is taken over all density/current profiles $( q_t,\rho_t)_{t \leq T}$ satisfying the conservation law \eqref{eq: conservation law} and the constraints at the boundaries of  each reservoir :
\begin{align}
\forall i \leq k, \qquad \frac{1}{T} \int_0^T ds \int_{\partial \bbD_i}  dx \; q(s,x) = u_i.
\end{align}
If the optimal $q(s,x)$ and $\rho(s,x)$ in  
(\ref{eq: time dependent LD}) are time independent, 
 this reduces to \emph{the additivity principle}, i.e. the following variational principle
\begin{align}
\label{eq: time independent LD}
\cG_{\bbD} (u) = \inf_{\rho} \Bigg \{  \int_\bbD dx 
\frac{ \big( q(x) + D \big( \rho(x) \big) \nabla \rho(x) \big)^2 }{2 \, \sigma \big( \rho(x) \big)} 
\Bigg\},
\end{align}
where the current is now in a stationary regime and thus time independent 
\begin{align}
\forall i \leq k, \qquad   \int_{\partial \bbD_i}  dx \; q(x) = u_i.
\end{align} 
To our knowledge, the  large deviation principle \eqref{eq: time dependent LD} hasn't been derived rigorously at this level of generality, nor the precise conditions allowing the reduction of the variational principle to the time independent version \eqref{eq: time independent LD}.
Nevertheless, for the SSEP it has been proven in \cite{BDGJL-2007-Theory of Probability,Bodineau-Lagouge-2012} that \eqref{eq: time independent LD} holds for simple geometries (periodic or $d=1$),   and  this corresponds to the case  $D (\rho)=1$ and $\sigma(\rho) = 2 \rho (1-\rho)$.

\subsubsection{Universality of the current fluctuations}

If the additivity principle \eqref{eq: time independent LD} holds then the variational problem  can be solved explicitly in one dimension.
This leads to concrete predictions on the cumulants of the current in one dimension \cite{Bodineau-Derrida-2004-PRL}.
In higher dimensions and for arbitrary reservoirs, the variational principle \eqref{eq: time independent LD} is much less explicit. Nevertheless, if there are only two types of reservoirs at densities $\rho_1,\rho_2$ then it has been shown in \cite{Akkermans-Bodineau-Derrida-Shpielberg} that for arbitrary geometries $\bbD$ the variational principle \eqref{eq: time independent LD} reduces to the one-dimensional functional  $\cG_{[0,1]}$ 
 on the domain $[0,1]$ with boundary conditions $\rho_1,\rho_2$ at 0 and 1 :
\begin{align}
\label{eq: universal 2 reservoirs}
\forall u \in \bbR, \qquad 
 \cG_{\bbD} (u) =    C(\bbD)  \cG_{[0,1]} \Big( \frac{u}{C(\bbD)}  \Big),
\end{align}
where $u$ is now a scalar associated with the total current flowing from reservoir $\partial \bbD_1$ to $\partial \bbD_2$ and the constant $C(\bbD)$ is the capacity of the domain $\bbD$ defined as 
\begin{align}
\label{eq: capacity C(D)}
C(\bbD)  = \int_\bbD |\nabla \Psi (x)|^2 dx \qquad 
\text{with} \qquad 
\begin{cases}
\Delta \Psi (x) = 0, \qquad x \in \bbD,\\
\Psi (x) = 0, \qquad x \in \partial \bbD_1,\\ 
\Psi (x) = 1, \qquad x \in \partial \bbD_2,
\end{cases} 
\end{align}
thus the constant $C(\bbD)$ depends only on the geometry of the domain $\bbD$.
The identity \eqref{eq: universal 2 reservoirs} can be understood easily as follows. 
Denoting by   $\bar \rho :[0,1] \mapsto \bbR^+$ the density of the steady state in  one dimension :
\begin{equation}
\forall r \in [0,1], \qquad \partial_r  \big( D( \bar \rho(r)) \partial_r \bar \rho(r) \big) = 0, 
\qquad \bar \rho(0) = \rho_1, \quad  \bar \rho(1) = \rho_2,
\end{equation}
then the density of the steady state in $\bbD$ is simply given by 
\begin{equation}
\label{eq: correspondence stationary}
\forall x \in \bbD, \quad 
\rho (x) = \bar \rho \big( \Psi(x)  \big)
\quad \text{as it solves} \quad \forall x \in \bbD, \quad \nabla   \big( D(  \rho(x)) \nabla  \rho (x) \big) = 0,
\end{equation}
with the boundary conditions 
 $\rho_1, \rho_2$ on $\partial \bbD_1, \partial \bbD_2$.
In particular, this shows that the steady state current is given by $\bar q(x) = \bar q^{(1)} \nabla \Psi(x)$ where 
  $\bar q^{(1)}  =- D(\bar \rho(u)) \bar \rho' (u)$  (for any  $u \in [0,1]$) stands for the constant steady state current on the interval $[0,1]$.  
{   The total current flowing through the second reservoir is obtained by integrating the current  along the normal direction on the boundary  $\partial \bbD_2$ in the   direction of the outward pointing normal $\vec{n}(x)$
\begin{align*}
\bar q = \int_{\partial \bbD_2}  \vec{n}(x) \cdot \bar q(x)
=  \bar q^{(1)}  \int_{\partial \bbD_1 \cup \partial \bbD_2} \Psi(x) \;  \vec{n}(x) \cdot \nabla \Psi(x)
=  \bar q^{(1)}  \int_\bbD  (\nabla \Psi(x))^2 = C(\bbD) \bar q^{(1)} ,
\end{align*}
where we first used that  $\Psi =0$ (respectively $\Psi =1$) on $\partial \bbD_1$ (respectively   on $\partial \bbD_2$)} and then the divergence Theorem to conclude by integration by parts.  The two steady state currents are therefore proportional up to a factor $C(\bbD)$, in agreement with formula \eqref{eq: universal 2 reservoirs}. 

This correspondence between the steady state profiles remains valid for the 
optimal profiles of the variational principle \eqref{eq: time independent LD}. Given $u$, if $\bar \rho^u : [0,1] \mapsto \bbR^+$ is a solution of the variational principle $\cG_{[0,1]}(u)$ in one dimension, then 
\begin{equation}
\label{eq: correspondence}
\forall x \in \bbD, \quad 
\rho^u (x) = \bar \rho^u \big( \Psi(x)  \big)
\end{equation}
is a solution for the problem in $\bbD$. 
We refer to \cite{Akkermans-Bodineau-Derrida-Shpielberg} for a derivation of \eqref{eq: universal 2 reservoirs}
from the mapping \eqref{eq: correspondence}.

The identity \eqref{eq: universal 2 reservoirs} implies that the cumulants of the current in the general domain $\bbD$ are proportional to the cumulants in $[0,1]$ which can be exactly computed. 
Note that \eqref{eq: universal 2 reservoirs} still holds if the reservoir $\partial \bbD_1$ (resp $\partial \bbD_2$) is made of several disconnected sets provided the density is equal to $\rho_1$ (resp $\rho_2$) on each type of reservoirs  \cite{Akkermans-Bodineau-Derrida-Shpielberg}.
This universality result confirms the numerical studies on fractal domains \cite{Groth-Tworzydlo-Beenakker-2008} and in two dimensions \cite{Akkermans-Bodineau-Derrida-Shpielberg}.
It was noticed in \cite{Akkermans-Bodineau-Derrida-Shpielberg} that in dimension 3, the cumulants are sensitive 
to the geometry of the reservoirs for microscopic reservoirs. 
This can already be seen at the level of the mean current in dimensions $d \geq 3$ which depends strongly on the local geometry \eqref{eq: mean current 3D}. 
In the following, we consider $k$ mesoscopic reservoirs, i.e. reservoirs with intermediate sizes $LR_i$ (at the microscopic scale) with  $1/L \ll R_i \ll 1$. 
For these mesoscopic reservoirs, \emph{we will assume that the macroscopic predictions of the additivity principle\eqref{eq: time independent LD}  remain valid  in the limits $\tau \to \infty$ first and then $L \to \infty$}.
We will argue that in this regime and in dimensions $d \geq 3$, the reservoirs become independent leading to simple expressions for the current large deviations.


\subsubsection{The case of two mesoscopic reservoirs}

We start by considering a domain $\bbD \subset [0,1]^d$ with 2 mesoscopic spherical sources $\bbB_1, \bbB_2$ of radii $R_1 , R_2 \ll 1$ centered at $x_1,x_2$ which are far apart  ($|x_1-x_2| \gg R_1,R_2$).
In this regime, sources are distant enough and the solution of the Laplace equation \eqref{eq: capacity C(D)}
can be approximated by adding the contributions of two independent charges 
\begin{equation}
\label{eq: fonction Psi}
\Psi(x) \simeq \frac{R_1^{d-2} R_2^{d-2}}{ R_1^{d-2} + R_2^{d-2}} \Big( \gp (x - x_2) - \gp (x - x_1) + \frac{1}{R_1^{d-2}} \Big),
\end{equation}
where $\gp (x) = \frac{1}{|x|^{d-2}}$ is the solution of the Laplace equation for a single charge in $\bbR^d$ with $d \geq 3$. 
For mesoscopic reservoirs $1/L \ll R_2 , R_1 \ll 1$, the function $\Psi$ varies only in the neighborhood of the reservoirs and  is essentially flat  elsewhere,  taking the constant value $\frac{R_2^{d-2}}{ R_1^{d-2} + R_2^{d-2}}$.
Recalling the correspondence \eqref{eq: correspondence}, the optimal profile associated with a deviation $u$  behaves as $\bar \rho^u \big( \Psi(x) \big)$ and varies only in the vicinity of the reservoirs. 
{
Denoting by $S_d$ the surface of the $d$-dimensional sphere, the capacity of the domain scales like 
\begin{align}
C(\bbD) &= \int_\bbD dx | \nabla \Psi|^2 \simeq 
S_d \, \left( \frac{R_1^{d-2} R_2^{d-2}}{ R_1^{d-2} + R_2^{d-2}} \right)^2  \left( \int_{R_1}^1 dr \frac{1}{(r^{d-1})^2} \, r^{d-1} + \int_{R_2}^1 dr \frac{1}{(r^{d-1})^2} \, r^{d-1}
\right) \nonumber \\
& = \frac{S_d}{d-2} \, \left( \frac{R_1^{d-2} R_2^{d-2}}{ R_1^{d-2} + R_2^{d-2}} \right)^2  
\left( \frac{1}{R_1^{d-2}} + \frac{1}{R_2^{d-2}} \right)  \simeq \frac{S_d}{d-2} \,\frac{R_1^{d-2} R_2^{d-2}}{ R_1^{d-2} + R_2^{d-2}}.
\label{eq: capacity}
\end{align}

Let us first consider the case of non symmetric reservoirs, say $R_2 \ll R_1 \ll 1$.
In this case the optimal profile density is mainly equal to the density $\rho_1$ of the  stronger reservoir and
the density varies only  close to the weaker reservoir $\partial \bbD_2$ which limits the current.
The capacity  scales like 
\begin{align}
C(\bbD) 
& \simeq \frac{S_d}{d-2} \, R_2^{d-2}. 
\label{eq: capacity single}
\end{align}
Applying the large deviations \eqref{eq: probability LD} and the identity \eqref{eq: universal 2 reservoirs} to reduce the computation to  one dimension leads to
\begin{align}
\label{eq: probability LD 2 mesoscopic 0}
\bbP \Big[ \frac{Q_{\tau}}{L^d \tau} = u \Big] 
\simeq \exp \Big( -  L^d \tau \, C(\bbD) \,  \cG_{[0,1]} \Big( \frac{u}{C(\bbD)}  \Big) \Big) .
\end{align}
Switching to the microscopic time units with $t = L^2 \tau$, we get  using the scaling \eqref{eq: capacity single} of the capacity for $1/L \ll R_2 \ll R_1 \ll 1$ that in the limit first $t \to \infty$ and then $L\to \infty$
\begin{align}
\label{eq: probability LD single mesoscopic}
\bbP \Big[ \frac{Q_t}{ L^{d-2} t} =  u \Big] 
\simeq \exp \Big( - L^{d-2} t  \;  \hat \cG^{\rho_1, \rho_2} (u)  \Big) ,
\end{align}
where the large deviation functional can be expressed only in terms of the smallest mesoscopic reservoir and the bulk density $\rho_1$ (imposed by the larger reservoir) 
\begin{align}
\label{eq: fonctionnelle  LD single mesoscopic}
\hat \cG^{\rho_1, \rho_2} (u) 
=  \frac{S_d R_2^{d-2}}{d-2}  \;    \cG_{[0,1]}^{\rho_1, \rho_2} \Big(  \frac{d-2}{S_d R_2^{d-2}}  u \Big)  .
\end{align}
The values of the boundary conditions have been added in upper-script 
{ (in $[0,1]$, the density at $0$ is $\rho_1$ and $\rho_2$ at 1).}
Note that  the formula is relevant only for $u$ of order $R_2^{d-2}$.  We stress that this functional records only the large deviations of the small reservoir and the bulk density. The rest of the geometry of the domain is  negligeable when $R_2 \ll R_1$.

We turn now to the case of   two mesoscopic reservoirs of comparable sizes $R_2 = c R_1 \ll 1$ (for some constant $c$).
The identity \eqref{eq: probability LD 2 mesoscopic 0} remains true and we are going to rewrite it below in order to isolate the contribution of each reservoir. For this we will use the additivity principle which says that for any $r \in [0,1]$
\begin{align}
\label{eq: additivity principle}
\cG_{[0,1]}^{\rho_1, \rho_2}  ( u ) 
 = \inf_{\hat \rho} \left\{ \cG_{[0,r]}^{\rho_1, \hat \rho}  ( u )   +
\cG_{[r,1]}^{\hat \rho, \rho_2}  ( u )  \right\} .
\end{align}
We will also use the following scaling identity of the one-dimensional large deviation function
\begin{align}
\label{eq: scaling LD 1d}
\cG_{[0,r]}^{\rho_1, \rho_2}  ( u ) = \frac{1}{r} \cG_{[0,1]}^{\rho_1, \rho_2}  ( r u ). 
\end{align}

According to the mapping \eqref{eq: correspondence}, the bulk density is equal to  
$\bar \rho^u  ( a )$ with {  $a = \frac{R_2^{d-2}}{ R_1^{d-2} + R_2^{d-2}}$,} 
as $\Psi (x)$ is essentially  equal to $a$  for $x$ away from $x_1, x_2$ in \eqref{eq: fonction Psi}. 
Using successively \eqref{eq: additivity principle} and \eqref{eq: scaling LD 1d}, the large deviation cost is therefore determined by
\begin{align*}
\cG_{[0,1]}^{\rho_1, \rho_2}  \Big( \frac{u}{C(\bbD)}  \Big) 
& = \cG_{[0,a]}^{\rho_1, \bar \rho^u  ( a )}  \Big( \frac{u}{C(\bbD)}  \Big)  +
\cG_{[a,1]}^{\bar \rho^u  ( a ), \rho_2}  \Big( \frac{u}{C(\bbD)}  \Big) \\
& = \inf_{\hat \rho} \left\{ \cG_{[0,a]}^{\rho_1, \hat \rho}  \Big( \frac{u}{C(\bbD)}  \Big)  +
\cG_{[a,1]}^{\hat \rho, \rho_2}  \Big( \frac{u}{C(\bbD)}  \Big) \right\}\\
& = \inf_{\hat \rho} \left\{ \frac{1}{a}  \cG_{[0,1]}^{\rho_1, \hat \rho}  \Big( a \frac{u}{C(\bbD)}  \Big)  +
\frac{1}{(1-a)} \cG_{[0,1]}^{\hat \rho, \rho_2}  \Big( (1-a)  \frac{u}{C(\bbD)}  \Big) \right\}.
\end{align*}
Using the value of $C(\bbD)$ in \eqref{eq: capacity}, we get 
$$
{ 
\frac{C(\bbD)}{a} =\frac{S_d}{d-2} R_1^{d-2}, \qquad  
\frac{C(\bbD)}{1-a} = \frac{S_d}{d-2} R_2^{d-2}.}
 $$
Thus we deduce from the previous computation that 
\begin{align}
C(\bbD) \cG_{[0,1]}^{\rho_1, \rho_2}  \Big( \frac{u}{C(\bbD)}  \Big) 
& = \inf_{\hat \rho} \left\{ 
\frac{S_d R_1^{d-2} }{d-2} \; \cG_{[0,1]}^{\rho_1, \hat \rho}  \Big(\frac{d-2}{S_d \, R_1^{d-2}}  u  \Big) +
\frac{S_d R_2^{d-2} }{d-2} \; \cG_{[0,1]}^{\rho_2, \hat \rho}  \Big( \frac{d-2}{S_d \, R_2^{d-2}}  u \Big)  
\right\} \nonumber\\
& = \inf_{\hat \rho} \left\{  \hat \cG^{\rho_1, \hat \rho} (u) 
+ \hat \cG^{\rho_2, \hat \rho} (u) \right\} .
\label{eq: additivity 2 reservoirs}
\end{align}
This shows that the large deviations associated with two mesoscopic reservoirs (of comparable sizes) is the superposition of the costs  associated to  each reservoir \eqref{eq: fonctionnelle  LD single mesoscopic}, up to optimising over the bulk density.

\begin{figure}[htbp]
\begin{center}
\includegraphics[width=3in]{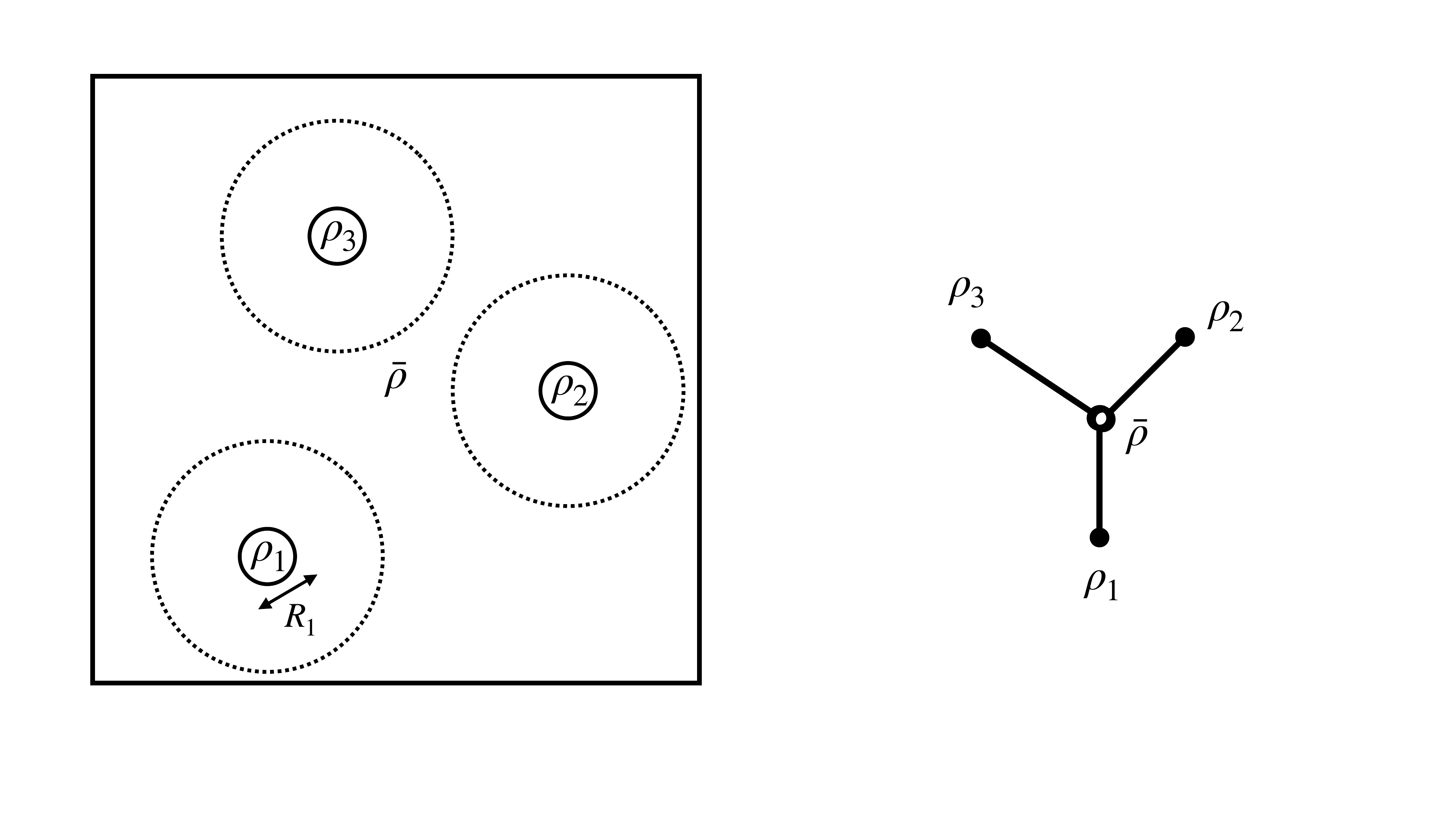} 
\caption{\small  Three mesoscopic reservoirs with densities $\rho_1, \rho_2, \rho_3$ are depicted, the variations of the density are located in disjoint small circles around the reservoirs (dashed lines). In the rest of bulk, the density is essentially constant, equal to $\bar \rho$.
For $t$ and $L$ large, the 3 reservoirs act as independent reservoirs in contact with the same bulk whose density $\bar \rho$ has to be optimised. The large deviation functional is no longer mapped to the unit segment but to a star shaped geometry. }
\label{fig: multiple mesoscopic sources}
\end{center}
\end{figure}

\subsubsection{Extension of the additivity principle}
\label{subsec: Extension of the additivity principle}

One  can  consider in the same way  $k$ mesoscopic reservoirs with densities $\rho_1, \dots, \rho_k$ in $d \geq 3$ (see Figure \ref{fig: multiple mesoscopic sources}). We suppose that the reservoir sizes $R_1, \dots , R_k$ are 
 comparable.
Let $u = \big( u_1, \dots, u_k \big)$ be the  fluxes flowing from each reservoir to the bulk such that $\sum_{i =1}^k u_i =0$.
We conjecture that  \eqref{eq: additivity 2 reservoirs} generalizes to
\begin{align}
\label{eq: probability LD k mesoscopic}
\bbP \Big[ \forall i \leq k, \quad  \frac{Q^{(i)}_t}{(R_i L)^{d-2} t} = u_i \Big] 
\simeq \exp \Big( -  L^{d-2} t \;   \inf_{\hat \rho} \left\{  \sum_{i =1}^k \hat \cG^{\rho_i, \hat \rho} (u_i) 
\right\}  \Big).
\end{align}
Thus the $k$ reservoirs become independent in the limit, except that they remain coupled through the bulk density (see Figure \ref{fig: multiple mesoscopic sources}): once a particle is ejected far from a reservoir it mixes fast and the detail of the bulk geometry is irrelevant.
The large deviation cost associated with the $i^{th}$-reservoir can be computed in a simple geometry as if the  
 reservoir at density $\rho_i$ was at the center of a sphere (of radius arbitrarily large) where the boundary is in contact with a reservoir $\bar \rho$ representing an infinite system \eqref{eq: fonctionnelle  LD single mesoscopic}.
}

\section{Conclusion}

In the present work, we have shown in Section \ref{sec: Microscopic computations}, using explicit expressions of the average current of the SSEP, that for large systems, 
the current generated by point contacts with reservoirs depends on 
 the dimensionality  $d$ of the graph. For strong enough point contacts, the current scales like ${1 \over L}$   in $d=1$ and ${1 \over \log L}$ in $d=2$,   when the contacts are  at distance of order $L$. 
  On the other hand, in $d=3$ and above, for large $L$, the current has a finite limit. This difference is closely related to the fact that, for infinite graphs,  random walks are recurrent in $d=1$ and $d=2$ while they are transient in $d \ge 3$. In fact,   
in  $d  \ge 3$ all the fluctuations of the current are dominated by the immediate neighborhood of the contacts (in particular  the current is influenced by their microscopic distances to the boundaries) making fluctuating hydrodynamics or the macroscopic fluctuation theory inoperative to predict the cumulants or the large deviations of the current. 
 In this case, understanding the statistics of the current would require the { knowledge of the microscopic  
 correlations} (see e.g Appendix \ref{Appendix: variance})  which is currently missing.
 
Still if instead of point contacts, one uses mesoscopic contacts with the reservoirs as in Section \ref{sec: Higher cumulants and large deviations}, (i.e. contacts of size much larger than the  lattice spacing but much smaller than the system size), we argue that the macroscopic fluctuation theory can be applied and if the additivity principle holds,  a stronger form of the universality  predicted in  \cite{Akkermans-Bodineau-Derrida-Shpielberg} can be { extended to}  an arbitrary number of reservoirs.
It remains an open problem  to generalise the universality results in \cite{Akkermans-Bodineau-Derrida-Shpielberg} for multiple reservoirs of  macroscopic sizes.

\appendix
\section{Green functions}
\label{appendix: Green functions}

In this appendix, we give  explicit expressions, numerical estimates   and some asymptotics of the Green function solution of \eqref{eq: Green function}
\begin{equation}
\forall x \in \Lambda, \qquad
\sum_{y \, \sim \,  x} \big( G^{(c)}_y  - G^{(c)}_x\big)   +   \delta_{x,c} = \frac{1}{|\Lambda|},
\label{eq: Green function-bis}
\end{equation}
in the case of an hypercubic lattice $\Lambda=\{1,2, \cdots ,L \}^d$ in dimension $d$ both in the case of periodic and  free boundary conditions.

\subsection{Periodic boundary conditions}
If  $x$ is a point of the  cube $\Lambda=\{1,2, \cdots ,L \}^d$, periodic boundary conditions means that 
$$G^{(c)}_x= 
G^{(c)}_{x+ L \, e_1} \cdots  
=
G^{(c)}_{x+ L \, e_d},
$$ 
where $e_i$ is the unit vector in direction $i$.
A solution of (\ref{eq: Green function-bis})
 (remember that one can  add an  arbitrary additive constant to the solution) 
is  obtained in terms of Fourier modes
\begin{equation}
G^{(c)}_x=
{1\over L^d} \sum_{k \ne (0,0 \cdots 0)} \
 {1\over 2 (d-\cos{ 2 \pi k_1 \over L} - \cdots \cos {2 \pi k_d \over L})}
   e^{ {2 i \pi  k.(x-c) \over L}}  ,
\label{G-per}
\end{equation}
where   the sum is over 
$k=(k_1,k_2, \cdots k_d)$ with $k_i \in \{0,1,\cdots, L-1 \}$ and $k$ non-zero: 
\\ \ 
\begin{itemize}
\item
{\bf In dimension 1 :} the Green function \eqref{G-per} is explicit
\begin{equation}
\label{G-per-1}
\begin{aligned}
G_x^{(c)}= {x^2 \over 2 L} - {(2 c-L)x \over 2 L}  
+ {L^2 - 6 c L +6 c^2 -1 \over 12 L} 
 \ \ \ \ \text{for} \ \ \ 1 \le x \le  c ,
\\
G_x^{(c)}= {x^2 \over 2 L} - {(2 c+L)x \over 2 L}  + {L^2 + 6 c L +6 c^2 -1 \over 12 L}
  \ \ \ \ \text{for} 
 \ \ \  c \le x \le L .
\end{aligned}
\end{equation}
\ \\ \
\item
{\bf In dimension 2 : }
Given a site $c$, we define 
 \begin{equation}
 c'= c+ {L e_1 \over 2}  \qquad \text{and} \qquad  c''=c + {L (e_1+e_2) \over 2}.
\label{eq: c' et c"}
\end{equation}
Then the Green function \eqref{G-per} takes the following values : 
\begin{center}
\begin{tabular}{ |c|c|c|c|c|} 
 \hline
\ \ \  \ \ $L$ \ \ \ \ \  & \ \ \ \ \  $G_c^{(c)} \ \ \ \ \  $ &\ \  $G_{c'}^{(c)}$ \ \  
& $G_{c''}^{(c)}$ & $2 \pi \, G_c^{(c)} / \log  L$ \\
 \hline
          2    &        0.156250    &       -0.031250    &       -0.093750    &        1.416363   \\
          3    &        0.222222    &       -0.000000    &       -0.055556    &        1.270934   \\
         400    &        1.002337    &       -0.027579    &       -0.055160    &        1.051140   \\
        1600    &        1.222972    &       -0.027579    &       -0.055159    &        1.041531   \\
        6400    &        1.443608    &       -0.027579    &       -0.055159    &        1.034961   \\
       25600    &        1.664244    &       -0.027579    &       -0.055159    &        1.030187   \\
     102400    &        1.884879    &       -0.027579    &       -0.055159    &        1.026559   \\
     409600    &        2.105515    &       -0.027580    &       -0.055159    &        1.023710   \\

 \hline
      $\infty$    &           &          &         &       1.   \\
 \hline
\end{tabular}
\end{center}

For large $L$, we see that for pairs of points at distances of order $L$, the Green function has a  finite limit, whereas for  points at distances of order 1, it grows logarithmically with $L$.
For example,  it is easy to show  from  (\ref{G-per}) that for large $L$
\begin{equation}
G_c^{(c)} \simeq {\log L  \over 2 \pi} ,
\quad G_{c'}^{(c)} = O(1), \quad G_{c''}^{(c)} = O(1),
\label{G-per-2}
\end{equation}
with the choices of $c',c''$ as in \eqref{eq: c' et c"}. The logarithmic growth of $G_c^{(c)}$ is obtained 
by integrating the singularity of the denominator in \eqref{G-per} for the modes close to 0.
Instead  { $G_{c'}^{(c)}$ and $G_{c''}^{(c)}$ have a finite limit due to 
 the fast oscillations of the numerator as $L$ goes to $\infty$.} 
\item {\bf  In dimension 3 :} 
Given a site $c$, we set 
 \begin{equation}
c'= c+ {L e_1 \over 2}  \qquad  \text{and} \qquad  c''=c + {L (e_1+e_2+e_3) \over 2} \ .  
\label{eq: c' et c"}
\end{equation}
Then the Green function \eqref{G-per} takes the following values :

\begin{center}
\begin{tabular}{ |c|c|c|c|c|c|} 
 \hline
\ \ \  \ \ $L$ \ \ \ \ \  &\ \ \ \ \  $G_{c}^{(c)}\ \ \ \ \  $ 
&\ \  $G_{c'}^{(c)}$ \ \  & $G_{c''}^{(c)}$ 
&\ \  $ L \, G_{c'}^{(c)}$ \ \  & $ L \, G_{c''}^{(c)}$  \\ 
 \hline

          2    &        0.151042    &        0.005208    &       -0.057292    &        0.010417    &       -0.114583   \\
         50    &        0.248216    &       -0.000150    &       -0.001278    &       -0.007521    &       -0.063896   \\
        100    &        0.250473    &       -0.000076    &       -0.000638    &       -0.007606    &       -0.063836   \\
        400    &        0.252167    &       -0.000019    &       -0.000160    &       -0.007632    &       -0.063817   \\
        800    &        0.252449    &       -0.000010    &       -0.000080    &       -0.007634    &       -0.063816   \\
       1600    &        0.252590    &       -0.000005    &       -0.000040    &       -0.007634    &       -0.063816   \\
       3200    &        0.252660    &       -0.000002    &       -0.000020    &       -0.007634    &       -0.063816   \\
       6400    &        0.252696    &       -0.000001    &       -0.000010    &       -0.007634    &       -0.063816   \\

 \hline
      $\infty$    &  0.252731          &          &         &     &      \\
 \hline
\end{tabular}
\end{center}
The extrapolated value is consistent with the known  value for lattice Green's function \cite{Guttmann} up to a simple  multiplicative factor $6$ due to  a difference in the  definition  of the Green function.

In dimension $3$, for large $L$, the Green function has a limit for pairs of points at distances of order $1$ while it decays like $1/L$ for pairs of points at distances of order $L$.
\end{itemize}

\subsection{Free boundary conditions}

For  free boundary conditions (sometimes called reflecting boundary conditions), 
if a site $x=(x_1,\cdots x_d)$  is at the boundary of the cube $\Lambda = \{1, \dots, L\}^d$,  i.e.  at least one of its coordinates $x_i=1$ or $L$, then the Green function has Neumann boundary conditions
\begin{align*}
G^{(c)}_x= 
G^{(c)}_{x-e_i}   \ \ \ \text{if} \ \ x_i=1 
 \ \ \ \ \ \ \ \  \text{and}  \ \ \  \ \ \ \ \ 
G^{(c)}_x= 
G^{(c)}_{x+e_i}   
  \ \ \ \text{if} \ \ x_i=L .
\end{align*}
A solution of 
(\ref{eq: Green function-bis}) with these boundary conditions is
\begin{equation}
G^{(c)}_x = 
{2^{d-1}\over L^d} \sum_{k \ne (0, \dots, 0)} \
\left(\prod_{i=1}^d 
{1\over \big(1 + \delta_{k_i,0}\big) }
\right)
{
\prod_{i=1}^d
\Big[ \cos \big(k_i(x_i-{1\over 2}) {\pi \over L} \big)
\cos\left(k_i(c_i-{1\over 2}) {\pi \over L}\right) \Big]
\, \over d- \cos\left({k_1 \pi \over L}\right)
-  \cdots 
- \cos\left({k_d \pi \over L}\right)}  ,
\label{G-open}
\end{equation} 
where   the sum is over the non-zero $k=(k_1,k_2, \cdots ,k_d)$ with $k_i \in \{0,1,\cdots, L-1 \}$.
\ \\ 
\begin{itemize}
\item {\bf In dimension 1 :}  the Green function \eqref{G-open} is explicit
\begin{equation}
\label{G-open-1}
\begin{aligned}
G_x^{(c)}= {x^2 \over 2 L} - {x \over 2 L}   + {2L^2 - 6 c L+3 L   +3  c^2 -3 c +1 \over 6 L} \ \ \ \ \text{for} \ \ \ 1 \le x \le  c, \\
G_x^{(c)}= {x^2 \over 2 L} - {(2 L+1 )x \over 2 L} + {2L^2 +3 L  +3 c^2 -3 c +1 \over 6 L}  \ \ \ \ \text{for} 
 \ \ \  c \le x \le L .
\end{aligned}
\end{equation} 
\ \\ \ 
\item {\bf  In dimension 2 :}
setting
\begin{equation}
c_1=(1,1), \qquad  c_2=(L,1), \qquad  c_3=(L,L ), \qquad c_4=(2,2)  , 
\end{equation}
then the Green function \eqref{G-open} takes the following values :
\begin{center} 
\begin{tabular}{ |c|c|c|c|c|c|}
 \hline
 $L$   &  $G_{c_1}^{(c_1)}  $
&\ \  $G_{c_2}^{(c_1)}$ \ \  
& $G_{c_3}^{(c_1)}$ &  
$ \pi \, G_{c_1}^{(c_1)}/ (2 \log L)$ 
& $ \pi \, G_{c_4}^{(c_4)}/(2 \log L)$ 
\\
\hline
          2    &        0.312500    &       -0.062500    &       -0.187500    &                0.708181    &        0.708181   \\
          3    &        0.541667    &       -0.083333    &       -0.208333    &                0.774475    &        0.317733   \\
          4    &        0.714286    &       -0.093750    &       -0.214286    &                0.809350    &        0.343974   \\
        200    &        3.191043    &       -0.110311    &       -0.220633    &                0.946051    &        0.788839   \\
        800    &        4.073580    &       -0.110317    &       -0.220635    &                0.957238    &        0.832619   \\
       3200    &        4.956122    &       -0.110318    &       -0.220636    &                0.964583    &        0.861368   \\
      12800    &        5.838665    &       -0.110318    &       -0.220636    &                0.969775    &        0.881690   \\
      51200    &        6.721207    &       -0.110318    &       -0.220636    &                0.973639    &        0.896815   \\
     204800    &        7.603750    &       -0.110318    &       -0.220636    &                0.976627    &        0.908512   \\
 \hline
      $\infty$    &   &           &   & 1 & 1    \\
 \hline
\end{tabular}
\end{center}

As in the case  of periodic boundary conditions,  for pairs of points at distances of order L the Green function has a finite limit, whereas for points at distances of order 1, it grows logarithmically with L.
Indeed with the previous notation, one can show from \eqref{G-open} the following asymptotics by integrating over the modes close to 0 :
\begin{equation}
G_{c_1}^{(c_1)} \simeq 
G_{c_4}^{(c_4)} \simeq {2 \, \log L  \over  \pi}, \quad 
\quad G_{c_2}^{(c_1)} = O(1), \quad G_{c_3}^{(c_1)} 
= O(1).
\label{G-open-2}
\end{equation}

\item {\bf In dimension 3 : }
setting 
\begin{align*}
c_1 &=(1,1,1), \qquad  c_2=(L,L,L), \qquad  c_3=(2,2,2), \\
c_4 &=(L-1,L-1,L-1), \qquad  c_5  =(1,L,1),
\end{align*}
one gets
\begin{center} 
\begin{tabular}{ |c|c|c|c|c|c|}
 \hline
$L$   &  $G_{c_1}^{(c_1)}  $
  & $G_{c_3}^{(c_3)}$
& $ L \, G_{c_2}^{(c_1)}$
&\ \  $ L \, G_{c_4}^{(c_3)}$ \ \  & $ L \, G_{c_5}^{(c_1)}$ 
\\  \hline  
          2    &        0.302083    &              0.302083    &             -0.229167    &       -0.229167    &        0.020833   \\
         50    &        0.702016        &        0.388686    &              -0.255244    &       -0.254443    &       -0.030324   \\
        100    &        0.711046    &               0.397701        &       -0.255259    &       -0.255059    &       -0.030483   \\
        200    &        0.715561    &             0.402215        &       -0.255263    &       -0.255213    &       -0.030523   \\
        400    &        0.717819      &        0.404472        &       -0.255264    &       -0.255251    &       -0.030533   \\
        800    &        0.718948    &              0.405601        &       -0.255264    &       -0.255261    &       -0.030535   \\
       1600    &        0.719512    &            0.406166        &       -0.255264    &       -0.255263    &       -0.030536   \\
       3200    &        0.719794    &               0.406448       &       -0.255264    &       -0.255264    &       -0.030536   \\
       6400    &        0.719936        &        0.406589       &       -0.255264    &       -0.255264    &       -0.030536   \\

 \hline
      $\infty$    &  0.720076     &          0.406731                   &  &  &    \\
 \hline
\end{tabular}
\end{center}

As for periodic boundary conditions, the Green function has a finite limit for pairs of points at a finite distance and it decays like $1/L$ for distances of oder $L$. A noticeable difference is that the Green function $G_c^{(c)}$ depends on the position of $c$ on the lattice. For example it is different for $c=(1,1,1)$ and $c=(2,2,2)$.

\end{itemize}

\section{Return probabilities}
\label{Appendix: return probability}

There are many analogies between the properties of  a random walk on a graph 
 and the current of the SSEP or of non  iteracting particles on the same graph.
In particular,  the return probability of a random walk is related to the Green function
 (see \cite{Montroll-Weiss-1965, Weiss-Rubin-1983, Doyle-Snell-1984, Wilmer-Levin-Peres})
as it satisfies equations similar to the steady state density \eqref{eq: Laplace 2 sites}.
 Let $P_x^{{a,b}}$ be the probability for a random walker, starting at position $x$ on a graph, to reach $a$ before ever visiting $b$. 
This probability  satisfies 
\begin{equation}
\sum_{y \, \sim \,  x} \big( P^{(a,b)}_y  - P^{(a,b)}_x\big)  +   K_0  \Big(\delta_{x,a} -\delta_{x,b}\Big) = 0, 
\label{P-eq}
\end{equation}
with the boundary conditions 
\begin{equation}
P^{(a,b)}_a=1  \qquad \text{and} \qquad  P^{(a,b)}_b=0.
\label{P-bound-cond}
\end{equation}
The solution is given in terms of the Green function  \eqref{eq: Green function}
$$
P^{a,b}_x= K_0 \Big(G_x^{(a)} -G_x^{(b)} \Big) + K_1,
$$
where the  two constants 
\begin{equation}
\label{eq: constants K0 K1}
K_0= \frac{1}{G_a^{(a)} + G_{b}^{(b)} -G_b^{(a)} -G_a^{(b)}} 
 \quad \text{and} \quad  
 K_1 = \Big( G_b^{(b)}- G_b^{(a)} \Big) \, K_0
 \end{equation}
  are determined by the boundary conditions (\ref{P-bound-cond}).
This leads to
\begin{equation}
\label{P-x}
P^{(a,b)}_x={ G_x^{(a)} - G_x^{(b)} + G_{b}^{(b)} -G_b^{(a)} 
\over  G_a^{(a)} + G_{b}^{(b)} -G_b^{(a)} -G_a^{(b)}}.
\end{equation}
Then the probability $r_a$ that a particle starting from $a$ returns to $a$ before visiting $b$ is
\begin{equation}
\label{eq: ra return}
r_a= 
{1 \over {\rm deg}(a)}
\sum_{y \, \sim \,  a}  P^{(a,b)}_y  = 1 - {K_0 \over  {\rm deg}(a)}= 1- 
{1 \over {\rm deg}(a)}
 \ { 1\over  G_a^{(a)} + G_{b}^{(b)} -G_b^{(a)} -G_a^{(b)}},
\end{equation}
where $ {\rm deg}(a)$ is is the number of sites connected to site $a$.
Comparing with (\ref{eq: solution J,K 2 sources})
gives a  link between the current $J$ of the SSEP and  the return probabilities $r_a$ or $r_b$.
In particular, the current $J$ vanishes  in the large $L$ limit when  the walk is recurrent, i.e. $r_a \to 1$.

\medskip

In Section \ref{sec: Simple random walks}, we discussed the case of a  random walk which is absorbed at rate $\gamma$ when it is on site $a$ and  at rate $\beta$ when it is on site $b$.
The probability $s_x$ that a walk starting in $x$ is finally  absorbed at site $a$ is solution of
\begin{align*}
\begin{cases}
\sum_{y \sim x} \big( s_y-s_x\big)  =0 \ \ \ \  \ \ &  \text{for} \  x \neq a \ \text{and}  \ x \neq b \\
\sum_{y \sim a} \big( s_y-s_a\big) + \gamma (1-s_a)=0 \ \ \ &  \text{for} \  x = a  \\
\sum_{y \sim b} \big( s_y-s_b\big) - \beta s_b=0 \  \ \ &  \text{for} \  x =  b
\end{cases}
\end{align*}
The solution is of the form $s_x = Q_0 P^{(a,b)}_x + Q_1$.
The parameters $Q_0, Q_1$ can be determined from the equations \eqref{P-eq}. 
Using notation \eqref{eq: constants K0 K1}, we get 
\begin{equation}
s_x={\gamma + \beta \gamma\,  \big(G_x^{(a)} -G_x^{(b)}- G_b^{(a)} +G_b^{(b)}\big) \over \beta+\gamma + \beta \gamma / K_0\,}
\quad \text{with} \quad 
K_0 =  {\rm deg}(a)\,  (1- r_a).
\end{equation}
where \eqref{eq: ra return} allows us to relate $K_0$ to $1-r_a$.
In particular, this gives
\begin{equation}
s_a=1- { \beta  \over \beta+\gamma + \beta \gamma /   [(1-r_a) \, {\rm deg}(a) ] }, 
\qquad 
s_b={ \gamma  \over \beta+\gamma + \beta \gamma /  [ (1-r_a)\,  {\rm deg}(a) ] }.
\label{random walk}
\end{equation}

%
%
%

\section{The variance}
\label{Appendix: variance}

To our knowledge, there does not exist an analytic expression of the cumulants of the current $Q_t$ for an arbitrary graph.  Already the calculation of the variance is difficult.
In this appendix, we recall how to obtain the variance of the current for the SSEP on an arbitrary graph in terms of the steady state two point correlations (see \cite[Section 3]{Bodineau-Derrida-Lebowitz-2008}).
We consider the SSEP on a general finite graph, in contact with two reservoirs at points $a$ and $b$ as in (\ref{eq: rates definition}). 
 Let us also introduce for each site $x$ of the graph a number $V_x$  solution for all $x \neq a,b$  of
\begin{equation}
\sum_{y \sim x} (V_y-V_x)=0 \label{V-eq}
\end{equation}
with the boundary conditions that $V_a=1$ and $V_b=0$.
Denoting by  $Q^{(x \to y)}_t$ the total integrated current  of particles from site $x$ to site $ y$, let us define 
 the weighted current $\widetilde{Q}_t$ as
$$
\widetilde{Q}_t = \sum_{x} \sum_{y \sim x} (V_x-V_y)  Q^{(x \to y)}_t.
$$
It is clear that every particle emitted in $a$ and absorded in $b$ contributes +1 to $ \widetilde{Q}_t $ while every particle emitted in $b$ and absorded in $a$ contributes $-1$ to this sum.
Thus at  leading order in $t$, it is equivalent to measure $\widetilde{Q}_t$ or the current  
$Q^{(a)}_t$ defined in \eqref{eq: microscopic current}.

The solution of (\ref{V-eq}) can be expressed in terms of Green functions \eqref{P-x}
\begin{equation}
\label{V-x-sol}
V_x={ G_x^{(a)} - G_x^{(b)} + G_{b}^{(b)} -G_b^{(a)}
\over  G_a^{(a)} + G_{b}^{(b)} -G_b^{(a)} -G_a^{(b)}}.
\end{equation}

\medskip

Let us now look at the evolution of  the first two moments of $\widetilde{Q}_t$.
Then 

\begin{align}
& {d \langle \widetilde{Q}_t \rangle \over dt} = \sum_{(x,y)} (V_x-V_y) \langle \eta_x -  \eta_y\rangle ,
\label{Qa}\\
& {d \langle \widetilde{Q}_t^2 \rangle \over dt} = 
\sum_{(x,y)} (V_x-V_y)^2 \, \langle \eta_x +  \eta_y - 2 \eta_x \eta_y\rangle 
+2  \sum_{(x,y)} (V_x-V_y) \,\Big\langle (\eta_x-\eta_y) \widetilde{Q}_t\Big\rangle ,
\label{Q2a}
\end{align}
where the sum is over all the edges $(x,y)$ on the graph. As the $(V_x)$ satisfy (\ref{V-eq}), we deduce from a discrete integration by parts that 
\begin{align}
& {d \langle \widetilde{Q}_t \rangle \over dt} =
 \sum_{y\sim a} (V_a-V_y) \,  \langle \eta_a \rangle 
 + \sum_{y\sim b} (V_b-V_y) \langle \eta_b \rangle ,
\label{Qb}\\
& {d \langle \widetilde{Q}_t^2 \rangle \over dt} = 
\sum_{(x,y)} (V_x-V_y)^2 \, \langle \eta_x +  \eta_y - 2 \eta_x \eta_y\rangle 
+2  \sum_{y\sim a} (V_a-V_y) \, \Big\langle \eta_a \,  \widetilde{Q}_t\Big\rangle 
+2  \sum_{y\sim b} (V_b-V_y) \, \Big\langle \eta_b\, \widetilde{Q}_t\Big\rangle .
\label{Q2b}
\end{align}
For simplicity, let us limit the discussion to the case of very strong contacts,  by taking the limit where rates at the contacts $\alpha,\beta,\gamma,\delta$ become very large keeping the ratios
${\alpha \over \gamma} = {\rho_a \over 1- \rho_a}$ and ${\delta \over \beta} = {\rho_b \over 1- \rho_b} $ fixed.
In this limit, the densities at $a,b$ become deterministic and decouple from the current  
\begin{equation}
\label{eq: decoupling densities}
\langle \eta_a\rangle \to  \rho_a, \ \   \langle \eta_b\rangle \to\rho_b
\quad \text{and} \quad
\big\langle \eta_a \widetilde{Q}_t \big\rangle \to \, \langle \eta_a \rangle \, \big\langle  \widetilde{Q}_t \big\rangle
, \ \ 
\big\langle \eta_b \widetilde{Q}_t \big\rangle \to \, \langle \eta_b \rangle \, \big\langle  \widetilde{Q}_t \big\rangle.
\end{equation}

Using \eqref{Q2b} and \eqref{eq: decoupling densities},  this leads to the following evolution of the variance
\begin{equation}
{d \langle \widetilde{Q}_t^2 \rangle_c \over dt} = 
{d \over dt} \Big(  \langle \widetilde{Q}_t^2 \rangle  -  \langle \widetilde{Q}_t \rangle^2 \Big) 
= 
\sum_{(x,y)} (V_x-V_y)^2 \, \langle \eta_x +  \eta_y - 2 \eta_x \eta_y\rangle 
\label{Q2c}
\end{equation}
which is an expression requiring the knowledge of  the pair correlations $\langle \eta_x \eta_y \rangle$.

\end{document}